\documentclass[10pt,
               aps,
               pre,
               twocolumn,
               showpacs, 
               groupedaddress]{revtex4-2}
\usepackage[utf8]{inputenc}
\usepackage{graphicx}

\usepackage[dvipsnames,table]{xcolor}
\usepackage[caption=false]{subfig}
\usepackage{amsmath, amsfonts, amssymb}
\usepackage{hyperref}
\usepackage{tikz}
\makeatletter
\newcommand*{\rom}[1]{\expandafter\romannumeral #1}
\makeatother

\begin{document}

\title{Activity induced trapping in a saw-tooth ratchet potential}
\author{M Muhsin}
\affiliation{Department of Physics, University of Kerala, Kariavattom, Thiruvananthapuram-$695581$, India}

\author{M Sahoo}
\email{jolly.iopb@gmail.com}
\affiliation{Department of Physics, University of Kerala, Kariavattom, Thiruvananthapuram-$695581$, India}

\date{\today}

\begin{abstract}
We consider an inertial active Ornstein-Uhlenbeck particle self-propelling in a saw-tooth ratchet potential. Using the Langevin simulation and matrix continued fraction method, the particle transport, steady state diffusion, and coherence in transport are investigated throughout the ratchet. Spatial asymmetry is found to be the key criterion for the possibility of directed transport in the ratchet. Interestingly, the simulated particle trajectories and the corresponding position and velocity distribution functions reveal that the system passes through an activity-induced transition in the transport from the running phase to the locked phase with the self-propulsion/activity time of the dynamics. This is further corroborated by the mean square displacement (MSD) calculation. The MSD gets suppressed with increase in the persistence of activity in the medium and finally approaches zero for very large value of self propulsion time, reflecting a kind of trapping of the particle by the ratchet for longer persistent of activity in the medium.
The non-monotonic behaviour of the particle current and Peclet number with self propulsion time confirms that the particle transport and it's coherence can be enhanced or reduced by fine tuning the persistent duration of activity.  
Moreover, for an intermediate range of self-propulsion time in the dynamics as well as for an intermediate range of mass of the particle, even though the particle current shows a pronounced unusual maximum with mass, there is no enhancement in the Peclet number, instead the Peclet number decreases with mass, confirming the degradation of coherence in transport.
Finally, from the analytical calculations, it is observed that for a highly viscous medium, where the inertial influence is negligibly small, the particle current approaches the current in the over damped regime.
\end{abstract}

\maketitle

\section{INTRODUCTION}
Noise is omnipresent and is an indispensable part of nature, which plays an important role in the dynamics of systems operating at microscopic length scale\cite{einstein1906theorie, langevin1908theory}.
A system which generates an unidirectional net transport out of a noisy environment in the molecular level utilizing non-equilibrium condition and spatial (or temporal) asymmetry is referred as a Brownian ratchet or Brownian motor\cite{reimann_brownian_2002, dean_thermo_1997, magnasco_forced_1993}. In the recent literature, active ratchets are realized through the use of active matter systems consisting of self propelled units that can be biological or non biological in nature \cite{reichhardt2017ratchet,angelani_active_2011, galajda2007wall, kaiser2014transport, koumakis2013targeted, bricard2013emergence, kummel2013circular}. Active matter is a special class of condensed matter systems, in which the individual constituents are having the ability to self-propel on their own by consuming energy from the environment. Such self-propelled particles are known as active particles and they are inherently driven far away from equilibrium, violating the well known fluctuation dissipation relation\cite{bechinger2016active, Gompper2020roadmap, magistris2015intro}. Examples of such active matter systems range from the microscopic to macroscopic length scale such as unicellular organisms like motile bacteria\cite{berg1972chemo, corbyn2021stochastic}, self motile Janus particles\cite{howse2007self, mallory2018active}, micro and nanorobots\cite{scholz2018rotating, palagi2018bioinspired}, hexbugs\cite{dauchot2019dynamics}, flocking of birds\cite{cavagna2015flocking}, school of fishes\cite{jhawar2020noise}, etc. Different models are proposed to study the dynamics of active matter both in single particle and in the collective level, such as Active Brownian particle(ABP) model\cite{hagen2009NonGaussian, hagen2011brownian, kanaya2020steady, lowen2020inertial}, Active Ornstein-Uhlenbeck particle(AOUP) model\cite{lehle2018analyzing, bonilla2019active, martin2021statistical}, run and tumble particle (RTP) model\cite{cates2012diffusive, cates2013when}, etc. 

Active ratchets are experimentally realized to generate unidirectional transport even in the absence of an external bias unlike passive Brownian ratchets \cite{reichhardt2017ratchet, kaiser2014transport, koumakis2013targeted}. When self-propelled particles are placed in an asymmetric potential, the particles on average can travel to the gentler side of the potential giving rise to unidirectional transport with a non-zero net particle flux\cite{kummel2013circular, koumakis2013targeted, kaiser2014transport}. Recently, there is an immense interest in the study of active ratchets and it is a growing field of research because of its enormous applications in the fabrication of different types of nanorobots, artificial swimmers, and other self-driven systems\cite{bechinger2016active}. The rectification effect of active matter in a periodic structure was first observed for run and tumble bacteria moving through funnel-shaped barriers\cite{galajda2007wall}. Subsequently, the rectification effects in active matter are studied both theoretically and numerically for different types of systems such as self-propelled particles on asymmetric substrates, bacterial colony, dusty plasma\cite{kaiser2014transport, koumakis2013targeted, koumakis_directed_2014, he2020experimental}, etc. The simulation results for the dynamics of active Janus particle in an asymmetric channel confirms that the rectification can be orders of magnitude stronger than that of ordinary thermal ratchets\cite{ghosh2013self}. Potosky et. al. found that the spatially modulated self-propelled velocity can also induce directed transport \cite{pototsky2013rectification}. Similarly, Angelani and co-workers observed active ratchet effects for run and tumble particles in a piecewise ratchet potential\cite{angelani_active_2011}. Rectification of twitching bacteria through 2D narrow channel is studied numerically in Ref.~\cite{bisht_rectification_2020}. The rectification effect in asymmetric periodic structures is found to be a general feature of active matter\cite{potiguar2014self, wan2008rectification, mijalkov2013sorting, angelani2009self, pototsky2013rectification, mcdermott2016collective, sandor2017collective, lambert2010collective, drocco2012bidirectional}.

However, most of the active ratchet studies are based on the overdamped dynamics of self propelled particles, where inertial effects are ignored. But the over damped approximation in the dynamical behavior is not appropriate in many situations such as granular matter in a diluted medium with high Reynold number, self propelling microdiode, colloidal particles in air, Janus particles in dusty plasm, and so on\cite{nagai2015collective, sharma2015remote, ivlev2015statistical, jung1996regular}. Recently, the dynamics of inertial active Brownian particles in a sawtooth potential results current reversal for an intermediate range of viscosity in the medium \cite{ai_transport_2017}. Similarly, the rotational rectification is investigated in a ratchet gear powered by active particles in Refs.~\cite{xu2021rotation, hatatani2022reversed}. A common interesting phenomena observed in these inertial ratchet models is the reversal of particle current. Although transport of inertial active particles is discussed in these models, the transport coherence remains largely unexplored. Coherence in transport in a stochastic environment is an important factor for determining the reliability of transport. 

In this work, we focus on the inertial active motion of a particle following Ornstein-Uhlenbeck process in a sawtooth ratchet potential. The dynamics of a passive Brownian particle in a sinusoidal potential driven by an exponentially correlated noise and Gaussian thermal noise is already discussed in Refs.~\cite{lindner_inertia_1999, bartussek_precise_1996}. In these models, the dynamics is mapped to a thermal bath at a certain temperature. However, in our model we consider the dynamics of an active particle in contact with an athermal bath. We mainly analyze the particle transport which is characterized by average current, diffusion, and the coherence in transport. One of our interesting findings is that an inertial active Ornstein-Uhlenbeck particle while self-propelling in a sawteeth ratchet potential, eventually gets trapped by the ratchet with longer persistence of self-propulsion in the environment. Both the particle transport and the coherence in transport show nonmonotonic behaviour with the self-propulsion time of the dynamics. Surprisingly, the current reversal is not observed in our model unlike the previously discussed inertial active ratchets in Refs.~\cite{ai_transport_2017, xu2021rotation, reichhardt2017ratchet}.

\section{MODEL AND METHOD}\label{sec:model}
\begin{figure}[!ht]
	\centering
	\includegraphics[scale=0.939]{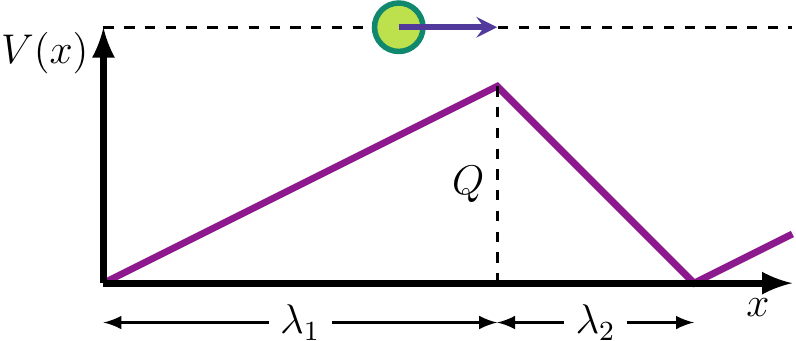}
	\caption{Schematic diagram of the ratchet potential [Eq.~\eqref{eq:ratchet}]}
	\label{fig:ratchet}
\end{figure}

We consider the motion of an active Ornstein-Uhlenbeck particle (AOUP) of mass $m$ through a ratchet potential. The dynamics of the particle is given by the Langevin's equation of motion\cite{lindner_inertia_1999,lowen2020inertial,arsha2021velocity, muhsin2021orbital, muhsin2022inertial} as
\begin{equation}
    m\ddot{x} = -\gamma \dot{x} - V^\prime(x) + \xi(t),
    \label{eq:model}
\end{equation}
with $x$ being the position co-ordinate and $v=\dot{x}$ as the velocity co-ordinate of the particle. Here, $\gamma$ is the viscous coefficient of the medium and $V(x)$ is the confining ratchet potential. $\xi(t)$ is the exponentially correlated noise with strength $C$, which follows the Ornstein-Uhlenbeck process\cite{bonilla2019active} as
\begin{equation}
    t_c \dot{\xi}(t) = -\xi(t) + \sqrt{2C}\ \eta(t).
    \label{eq:noise-dynamics}
\end{equation}
Here, $\eta(t)$ is the delta correlated Gaussian white noise which satisfies the properties $\langle \eta(t) \rangle = 0$ and $\langle \eta(t) \eta(s) \rangle = \delta(t - s)$. The angular bracket $\langle \cdots \rangle$ denotes the ensemble average over noise. The statistical properties of the Ornstein-Uhlenbeck (OU) noise $\xi(t)$ is given by
\begin{equation}
    \langle \xi(t) \rangle = 0,\quad \langle \xi(t) \xi(s) \rangle = \frac{C}{t_c}\exp\left(-\frac{|t - s|}{t_c}\right),
    \label{eq:noise-corr}
\end{equation}
with $t_{c}$ being the noise correlation time. It is the time up to which the particle self-propels in the ratchet and hence activity persists in the medium for a time interval of $t_{c}$. A finite $t_{c}$ notably quantifies the presence of activity or correlation in the medium, that decays exponentially with $t_{c}$. For a nonzero $t_{c}$ value, the system is inherently driven away from equilibrium\cite{tailleur2009sedimentation}. However, In the passive limit ($t_{c} \rightarrow 0$ limit) of our model, we consider the strength of noise $C$ to be $\gamma k_{B} T$ (fluctuation-dissipation relation) in order for the system to approach the typical thermal equilibrium limit of the dynamics at temperature $T$ \cite{fodor2016howfar, mandal2017entropy}. 
The potential $V(x)$ that appears in Eq.~\eqref{eq:model} has the form
\begin{equation}
    V(x) =  \begin{cases} 
      \dfrac{Q}{\lambda_1} x, & x\leq \lambda_1 \\[1.0em]
      \dfrac{Q}{\lambda_2} (\lambda-x), & \lambda_1 < x \leq \lambda.
  \end{cases}
  \label{eq:ratchet}
\end{equation}
Here, $Q$ is the potential height and $\lambda = \lambda_1 + \lambda_2$ is the periodicity of the ratchet potential (see Fig.~\ref{fig:ratchet}). Eq.~\eqref{eq:ratchet} represents a sawtooth potential which is symmetric when $\lambda_1 = \lambda_2$. Therefore, we introduce an asymmetric parameter $\Delta$ such that $\Delta = \lambda_1 - \lambda_2$. 

In this work, we are mainly interested in the particle transport, associated dispersive spread or diffusion, and coherence in transport of the particle. The particle transport can be quantified by measuring an essential quantity, known as particle current. As per the geometry of the ratchet potential, the motion of the particle is along $x$-direction and hence the average particle current in the stationary state can be defined as\cite{reimann_brownian_2002, ai2009directed},
\begin{equation}
    \langle j \rangle  = \lim_{t\rightarrow \infty} \left\langle \frac{x(t) - x(0)}{t} \right\rangle.
    \label{eq:current}
\end{equation}
Similarly, the diffusive spread or diffusion can be quantified by measuring the diffusion coefficient $D$ about the mean position of particle, which is given by \cite{lindner2008diffusion},
\begin{equation}
    D = \lim_{t\rightarrow \infty} \frac{\langle x^2 \rangle - \langle x \rangle^2}{2t}.
    \label{eq:diffusion}
\end{equation}
Here, $\langle x^2 \rangle - \langle x \rangle^2$ can be characterized as the mean square displacement (MSD) of the particle.
The transport of the particle in such an asymmetric potential and stochastic environment depends on the diffusive spread and the mean velocity of the particle. The effectiveness or coherence in the transport can be quantified by measuring a dimensionless parameter called Peclet number $Pe$, which is defined as 
\begin{equation}
    Pe = \frac{\langle j \rangle \lambda}{D}.
    \label{eq:Pe}
\end{equation}
We have set the periodicity of the potential $\lambda$ as unity throughout this paper.
\section{RESULTS AND DISCUSSION}\label{sec:result}
The Fokker-Planck equation corresponding to the dynamics in  Eq.~\eqref{eq:model} for the probability density function $P(x, v, \xi;t)$ is given by
\begin{equation}
\begin{split}
    \frac{\partial P}{\partial t} = -v \frac{\partial P}{\partial x} & + \frac{\partial}{\partial v}\left( \frac{\gamma v}{m} + \frac{V^\prime (x)}{m} - \frac{\xi(t)}{m} \right)P \\
    & + \frac{\partial }{\partial \xi} \left( \frac{\xi}{t_c} + \frac{C}{t_c^2}\frac{\partial}{\partial \xi} \right)P.
\end{split}
   \label{eq:FPE-under}  
\end{equation}
It is not possible to obtain the exact analytical solution of Eq.~\eqref{eq:FPE-under} even for the steady state because of the non-linearity in the  gradient of the potential function $V(x)$. However, the steady-state solution is possible for $P(x, v, \xi;t)$ with the help of various numerical approximation schemes. In order to investigate the transport of the particle, one can solve the dynamics either by directly simulating Eq.~\eqref{eq:model} or by employing the Matrix continued fraction method (MCFM) to Eq.~\eqref{eq:FPE-under} for the approximate steady state solution of $P(x, v, \xi)$.
\begin{figure*}[!ht]
    \centering
    \includegraphics[scale=0.673]{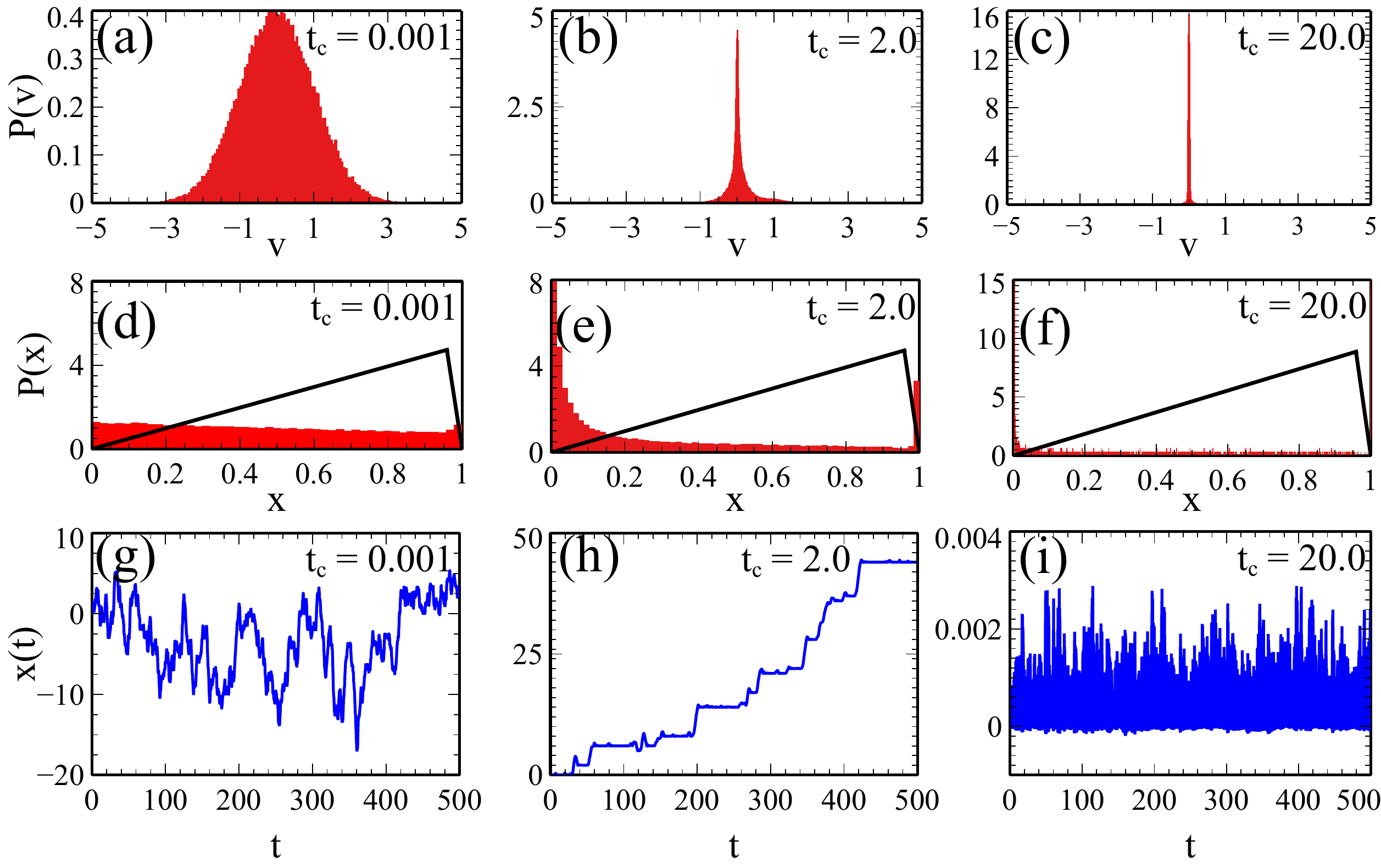}
    \caption{(a-c) Velocity distribution $P(v)$ as a function of $v$ for different values of $t_{c}$. (d-f) Position distribution $P(x)$ as a function of $x$ for different values of $t_{c}$. The black solid line represents the corresponding saw-tooth potential. (g-i) Particle trajectories are plotted for different values of $t_{c}$. The common parameters taken are $m = 1.0$, $\Delta = 0.9$, $Q = 0.5$, and $\gamma=1.0$.}
    \label{fig:dist-tc}
\end{figure*}

For the overdamped dynamics of the particle, the inertial term in Eq.~\eqref{eq:model} is neglected and the corresponding probability density function $P(x, \xi; t)$ satisfies the Fokker-Plank equation \cite{risken1996fokker, bartussek_precise_1996},
\begin{equation}
    \frac{\partial P}{\partial t} = \frac{\partial}{\partial x}\left( \frac{V'(x)}{\gamma} - \frac{\xi}{\gamma}\right)P + \frac{\partial}{\partial \xi} \left( \frac{\xi}{t_c} + \frac{C}{t_c}\frac{\partial}{\partial \xi} \right)P.
    \label{eq:FPE-over}
\end{equation}
In the stationary state or steady state limit, the probability density function $P(x, \xi;t)$ satisfies 
\begin{equation}
    \frac{\partial P}{\partial t} = 0.
    \label{eq:FPE-over-st}
\end{equation}
In order to find the approximate solution of Eq.~\eqref{eq:FPE-over-st} for the stationary state probability distribution function $P(x, \xi)$, it can be expanded in complete sets of functions in both variables $x$ and $\xi$ using set of Hermite functions. Since the potential is periodic in nature, $P(x,\xi$) can take the form~\cite{bartussek_precise_1996}
\begin{equation}
    P(x, \xi) = \phi_0(\xi) \sum_{p=0}^{\infty} \sum_{\mu=-\infty}^{\infty} c_{p}^{\mu} e^{2\pi i \mu x/\lambda} \phi_p(\xi).
    \label{eq:pdf-over}
\end{equation}
Here, the prefactor $\phi_0(\xi)$ is introduced for the simplification of mathematical calculations and $\phi_p(\xi)$ is the set of Hermite functions given by
\begin{equation}
    \phi_p(\xi) = \frac{1}{\sqrt{\alpha 2^pp!\sqrt{\pi}}}e^{\frac{-\xi^2}{2\alpha^2}}H_p\left(\frac{\xi}{\alpha}\right),
    \label{eq:phipx}
\end{equation}
with $\alpha$ as the scaling parameter considered as $\alpha = \sqrt{\frac{2D}{t_c}}$ and $H_p(x)$ is the Hermite polynomial. Since the potential $V(x)$ is periodic in nature, the force exerted by the potential, $f(x) = -V'(x)$ can be expanded in terms of Fourier series as
\begin{equation}
    f(x) = \sum_{l=-\infty}^{\infty} f_l e^{2\pi i l x/\lambda}.
    \label{eq:f-expansion}
\end{equation}
Substituting Eq.~\eqref{eq:pdf-over} in Eq.~\eqref{eq:FPE-over-st} and using Eq.~\eqref{eq:phipx} and Eq.~\eqref{eq:f-expansion},
we obtain a tridiagonal vector recurrence relation in terms of expansion coefficients $c_p^\mu$ as
\begin{equation}
    Q_p^- c_{p-1} + Q_p c_p + Q_p^+ c_{p+1} = 0,
    \label{eq:tridiag-over}
\end{equation}
with
\begin{align}
    Q_p^- &= \sqrt{\frac{pC}{t_c}} B,\\
    Q_p   &= A - \frac{p}{t_c} I, \\
    \text{and}\quad Q_p^+ &= \sqrt{\frac{(p+1)C}{t_c}} B.
\end{align}
\begin{figure}[!ht]
    \centering
    \includegraphics[scale=0.537]{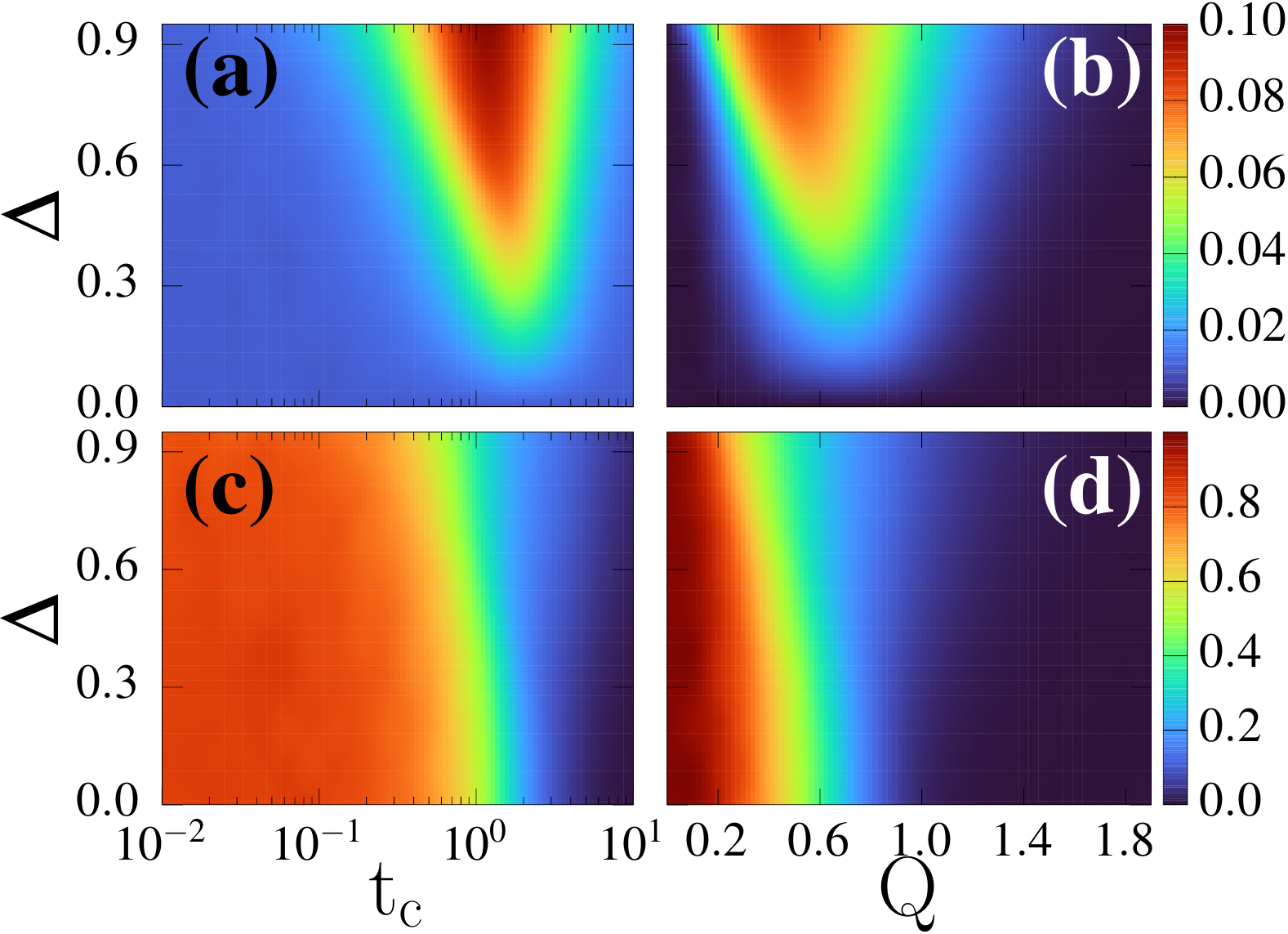}
    \caption{2D plot of $\langle j \rangle$ and $D$ as a function of $\Delta$ and $t_c$ in (a) and (c), respectively. 2D plot of $\langle j \rangle$ and $D$ as a function of $\Delta$ and $Q$ in (b) and (d), respectively. The common parameters in (a) and (c) are ($m = 1.0$,\ $Q = 0.5$) and in (b) and (d) are ($m = 1.0,\ t_c = 1.0$).}
    \label{fig:2d_2}
\end{figure}
Here, $c_p$ is a column vector consisting of the elements $c_p^0, \ c_p^1,\ c_p^2 \cdots c_p^\mu$ . The elements of matrices $A$ and $B$ are given by
\begin{align}
    \left[ A_{n,m} \right] &= \frac{2\pi in}{\gamma \lambda} f_{n - m} ,\\
    \left[ B_{n,m} \right] &= -\frac{2\pi im}{\gamma \lambda} \delta_{n,m},
\end{align}
with $I$ being the identity matrix. The vector recurrence relation in Eq.~\eqref{eq:tridiag-over} can be solved numerically using the matrix continued fraction method as described in Ref. \cite{risken1996fokker}. For this purpose, we introduce the matrix $S_p$ such that
\begin{equation}
    c_{p+1} = S_p c_p.
    \label{eq:Sp}
\end{equation}
Now substituting Eq.~\eqref{eq:Sp} in Eq.~\eqref{eq:tridiag-over}, we obtain 
\begin{equation}
    Q_p^- c_{p-1} + \left( Q_p + Q_p^+S_p \right)c_p = 0.
    \label{eq:intermediate}
\end{equation}
Further solving Eq.~\eqref{eq:intermediate}, we obtain the matrix $S_{p}$ as the matrix continued fraction: 
\begin{equation}
    S_p = -\left( Q_{p+1} + Q_{p+1}^+S_{p+1} \right)^{-1} Q_{p+1}^-.
    \label{eq:mcf-over}
\end{equation}
For $p=0$, Eq.~\eqref{eq:intermediate} takes the form
\begin{equation}
    \left( Q_0 + Q_0^+S_0 \right)c_0 = 0.
    \label{eq:matrix-c0}
\end{equation}
Normalization of the steady state probability distribution $P(x,\xi)$
\begin{equation}
    \int\limits_0^\lambda dx \int\limits_{-\infty}^{\infty} d\xi\; P(x, \xi) = 1,
\end{equation}
yields 
\begin{equation}
    c_0^0 = \frac{1}{\lambda}.
\end{equation}
Using this arbitrary component $c_{0}^0$ in Eqs.~\eqref{eq:matrix-c0} and \eqref{eq:Sp}, one can compute all the components of $c_{p}$. In order to find the average particle current, the Fokker-Planck equation [Eq.~\eqref{eq:FPE-over}] can be written in the form of a continuity equation as
\begin{equation}
    \frac{\partial P(x, \xi; t)}{\partial t} = -\frac{\partial \rho_x(x, \xi; t)}{\partial x} -\frac{\partial \rho_\xi(x, \xi; t)}{\partial \xi},
    \label{eq:continuity}
\end{equation}
where $\rho_x(x, \xi; t)$ and $\rho_\xi(x, \xi; t)$ are the probability currents in the $x$ and $\xi$ directions, respectively. Now Comparing Eq.~\eqref{eq:continuity} with Eq.~\eqref{eq:FPE-over}, we have 
\begin{equation}
    \rho_x(x, \xi ; t) = \left( \frac{f(x)}{\gamma} - \frac{\xi}{\gamma} \right) P(x, \xi; t).
\end{equation}
Hence, the average stationary current in the $x$ direction over a period is given by
\begin{align}
    \langle j \rangle &= \frac{1}{\lambda} \int\limits_0^\lambda dx \int\limits_{-\infty}^{\infty} d\xi\; \rho_x^{(st)}(x, \xi) \nonumber \\
    &= \frac{1}{\gamma} \left[ -\sum_{\mu=-\infty}^{\infty} f_\mu c_0^{-\mu} + \frac{C}{t_c}c_1^0 \right].
    \label{eq:j_over_MCFM}
\end{align}
\begin{figure}[!ht]
    \centering
    \includegraphics[scale=0.3]{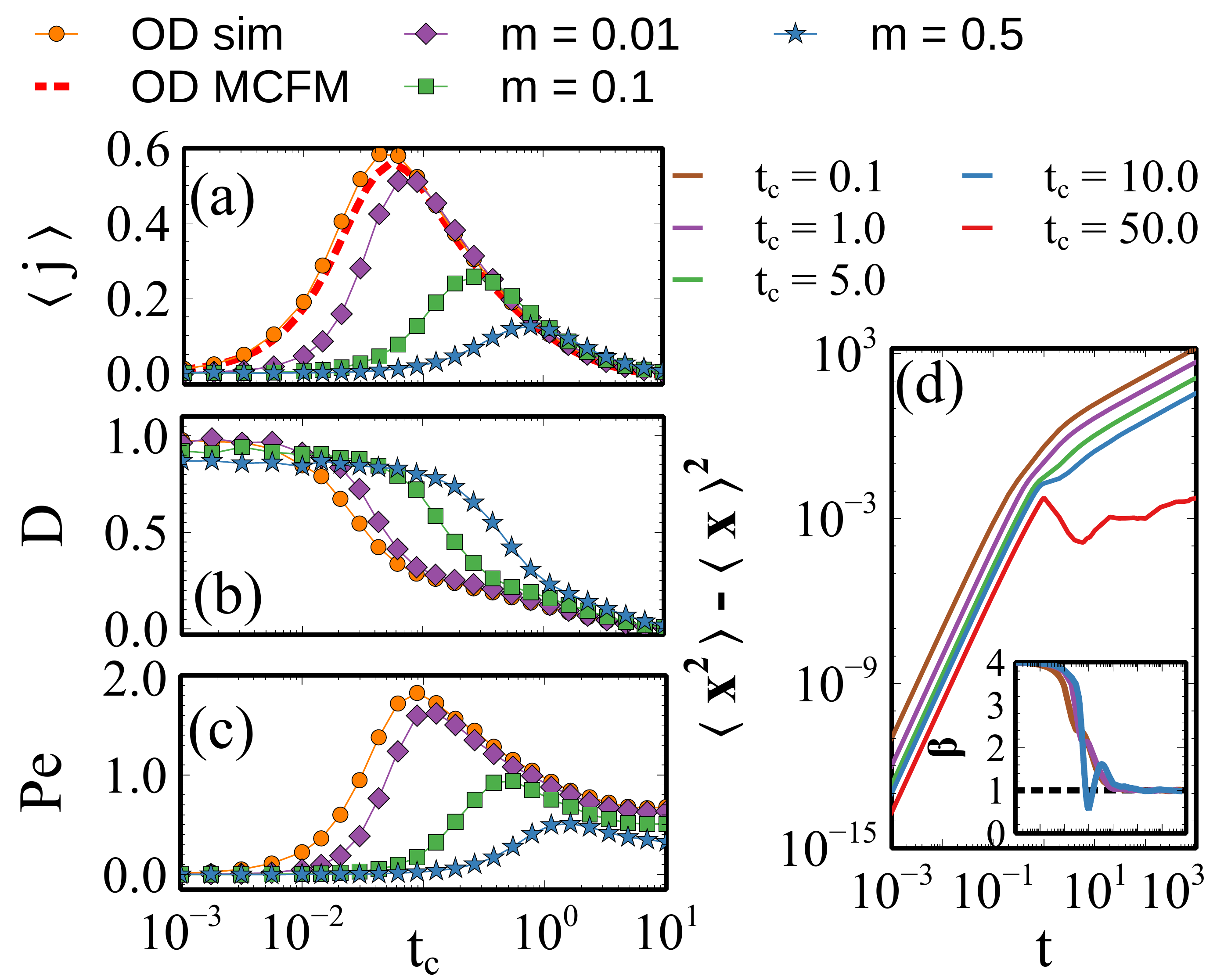}
    \caption{$\langle j \rangle$, $D$, and $Pe$ as a function of $t_c$ for different values of $m$ are shown in (a), (b), and (c), respectively. The results of simulation (OD sim) and MCFM calculation (OD MCFM) for overdamped case are also plotted as a function of $t_{c}$. MSD as a function of $t$ is shown in (d) for different values of $t_{c}$ and for $m=0.5$. Inset of (d) shows the exponent $\beta$ as a function of $t$. The other common parameters are $\Delta = 0.9$ and $Q = 0.5$.}
    \label{fig:tc_for_m}
\end{figure}
Proceeding in the same way as in overdamped case described above, one can solve Eq.~\eqref{eq:FPE-under} for the steady state probability distribution $P(x,v;\xi)$ and find out the average particle current. The approximate steady state solution of Eq.~\eqref{eq:FPE-under} can take the form
\begin{equation}
    P(x, v, \xi) = \phi_0(\xi) \psi_0(v) \sum_{r=0}^{\infty} \sum_{p=0}^{\infty} \sum_{\mu=-\infty}^{\infty} c_{p, r}^{\mu} e^{2\pi i \mu x/\lambda} \phi_p(\xi) \psi_r(v).
    \label{eq:pdf-under}
\end{equation}
Here, $\psi_r(v)$ is the Hermite function given by
\begin{equation}
    \psi_r(v) = \dfrac{1}{\sqrt{\beta 2^r r!\sqrt{\pi}}}e^{\frac{-v^2}{2\beta^2}}H_r\left(\frac{v}{\beta}\right),
    \label{eq:psirx}
\end{equation}
with $\beta$ being a scaling parameter. Following the same method discussed earlier, we get the recursion relation in $c_{p,r}^\mu$
\begin{equation}
\begin{split}
    A_{p,r}\ c_{p, r-2} + & B_{p,r}\ c_{p, r-1} + \Gamma_{p,r}\ c_{p,r} + E_{p,r}\ c_{p, r+1} \\
    & + Z_{p,r}\ c_{p-1, r-1} + \Theta_{p,r}\ c_{p+1,r-1} = 0.
\end{split}
\label{eq:recurr_under}
\end{equation}
Here, $A,\ B,\ \Gamma,\ E,\ Z$ and $\Theta$ are matrices whose elements are given by
\begin{align*}
    [A_{\mu,\nu}]_{p,r} &= -\frac{\gamma}{m} \sqrt{(r-1)r} \ \delta_{\mu,\nu},\\
    [B_{\mu,\nu}]_{p,r} &= \frac{\sqrt{2r}}{\beta m} f_{\mu - \nu} - \frac{i \nu k \beta\sqrt{r}}{\sqrt{2}}\ \delta_{\mu,\nu}, \\
    [\Gamma_{\mu, \nu}]_{p,r} &= -\left( \frac{\gamma r}{m} + \frac{p}{t_c} \right)\ \delta_{\mu,\nu}, \\
    [E_{\mu, \nu}]_{p,r} &= -\frac{i \nu k \beta \sqrt{r+1}}{\sqrt{2}}\ \delta_{\mu,\nu}, \\
    [Z_{\mu, \nu}]_{p,r} &= \frac{\alpha \sqrt{r p}}{m \beta}\ \delta_{\mu,\nu}, \\
    \text{and}\quad [\Theta_{\mu, \nu}]_{p,r} &= \frac{\alpha}{m \beta} \sqrt{r(p + 1)}\ \delta_{\mu,\nu},
\end{align*} respectively. $c_{p,r}$ is a column matrix given as
\begin{equation*}
    c_{p,r} = \left[\cdots\ \ c_{p,r}^{-1}\ \ c_{p,r}^0\ \ c_{p,r}^1 \ \ \cdots\ \  \right]^T.
\end{equation*}
Now, by solving Eq.~\eqref{eq:recurr_under} and using the column vector $c_{p,r}$, one can compute the steady state probability distribution and the average current. 

We have also simulated the dynamics [Eq.~\eqref{eq:model}] using Huen's method algorithm. The simulation was run for $10^5$ seconds and averaged over $10^4$ realization. The simulation results of trajectory and position and velocity distributions of the particle for different values of $t_c$ are shown in Fig.~\ref{fig:dist-tc}. For very small value of $t_c$, the particle is distributed uniformly throughout the potential as shown in Fig.~\ref{fig:dist-tc}(d). Thus for very small $t_c$, the particle is merely influenced by the presence of barriers exerted by the ratchet potential and hence distributed uniformly throughout the space. As a result, the velocity distribution is almost Gaussian [see Fig.~\ref{fig:dist-tc}(a)]. This can also be confirmed by looking at the simulated trajectory of the particle, which does not show any signature of the presence of potential trap. This behavior can be understood as follows. In the steady-state, the magnitude of the noise correlation of the OU process [Eq.~\eqref{eq:noise-corr}] varies inversely with the correlation time $t_c$, such that $\left\langle \xi^2(t) \right\rangle = \frac{C}{t_c}$. Hence, for a very small value of $t_c$, even though the noise correlation persists for a very small interval of time, the intensity of the correlation is very high. As a result, the magnitude of random kicks on the particle is very large. Due to this, the particle does not feel the presence of the potential barrier and moves freely both forward and backward directions of ratchet potential, resulting an uniform distribution of the particles in $t_{c} \rightarrow 0$ limit. In this limit, the system behaves as if it is in the running state.

With further increase in $t_{c}$, the magnitude of the noise correlation decreases and at the same time the duration of its persistence increases. The particle starts getting more and more confined at the potential minima and feels the influence of the barriers in both the forward and backward directions of the ratchet potential. This is very well reflected from the position distribution of the particle in Fig.~\ref{fig:dist-tc}(e) with maximum probability of finding the particle in one of the potential minima. Due to presence of asymmetry in the potential, the particle on an average makes more jumps towards the forward direction as compared to the backward direction of the potential. This can also be seen from the trajectory plotted in Fig.~\ref{fig:dist-tc}(h), where there are sudden jumps and stable regions indicating the presence of potential being felt by the particle.  As a result, the velocity distribution becomes non-Gaussian with exponential tails in both the directions [see Fig.~\ref{fig:dist-tc}(b)]. In this regime, the value of $\left\langle \xi^2(t) \right\rangle$ is such that the particle becomes capable of overcoming the potential barrier in the direction with gentler slope of the potential. Hence, on an average, a non-zero particle current is expected. For very large $t_{c}$, the magnitude of the noise correlation [Eq.~\eqref{eq:noise-corr}] becomes very small and the correlation persists for longer interval of time. Hence, the magnitude of the random kicks are very small and are not strong enough to make the particle escape from the potential minimum [see Fig.~\ref{fig:dist-tc}(f-i)]. This is the reason for which the velocity distribution approaches a delta function centered at zero [Fig.~\ref{fig:dist-tc}(c)] for very large $t_{c}$ value. In this limit, the system behaves as if it is trapped or in the locked state. Thus, with $t_{c}$ the system passes through a transition from the running state to the locked state of the particle transport.

Next, we have simulated the steady-state particle current $\langle j \rangle$ and diffusion coefficient $D$ for different values of $\Delta$, $t_c$, and $Q$, which are shown in Fig.~\ref{fig:2d_2}. The two dimensional (2D) plots of $\langle j\rangle$ and $D$ as a function of $\Delta$ and $t_{c}$ are shown in Fig.~\ref{fig:2d_2}(a) and (c), respectively. Similarly, we have shown the 2D plots of $\langle j \rangle$ and $D$ as a function of $\Delta$ and $Q$ in Fig.~\ref{fig:2d_2}(b) and (d), respectively. From these plots, it is observed that for a given spatial asymmetry in the potential, $\langle j \rangle$ shows a non- monotonic behavior with both $t_c$ and $Q$ values. Further, the maximum current is found to be sensitive to the spatial asymmetry of the potential and it increases with increase in the asymmetry parameter $\Delta$. Whereas diffusion shows decreasing behavior with both $t_{c}$ and $Q$.
\begin{figure}[!ht]
    \centering
    \includegraphics[scale=0.54]{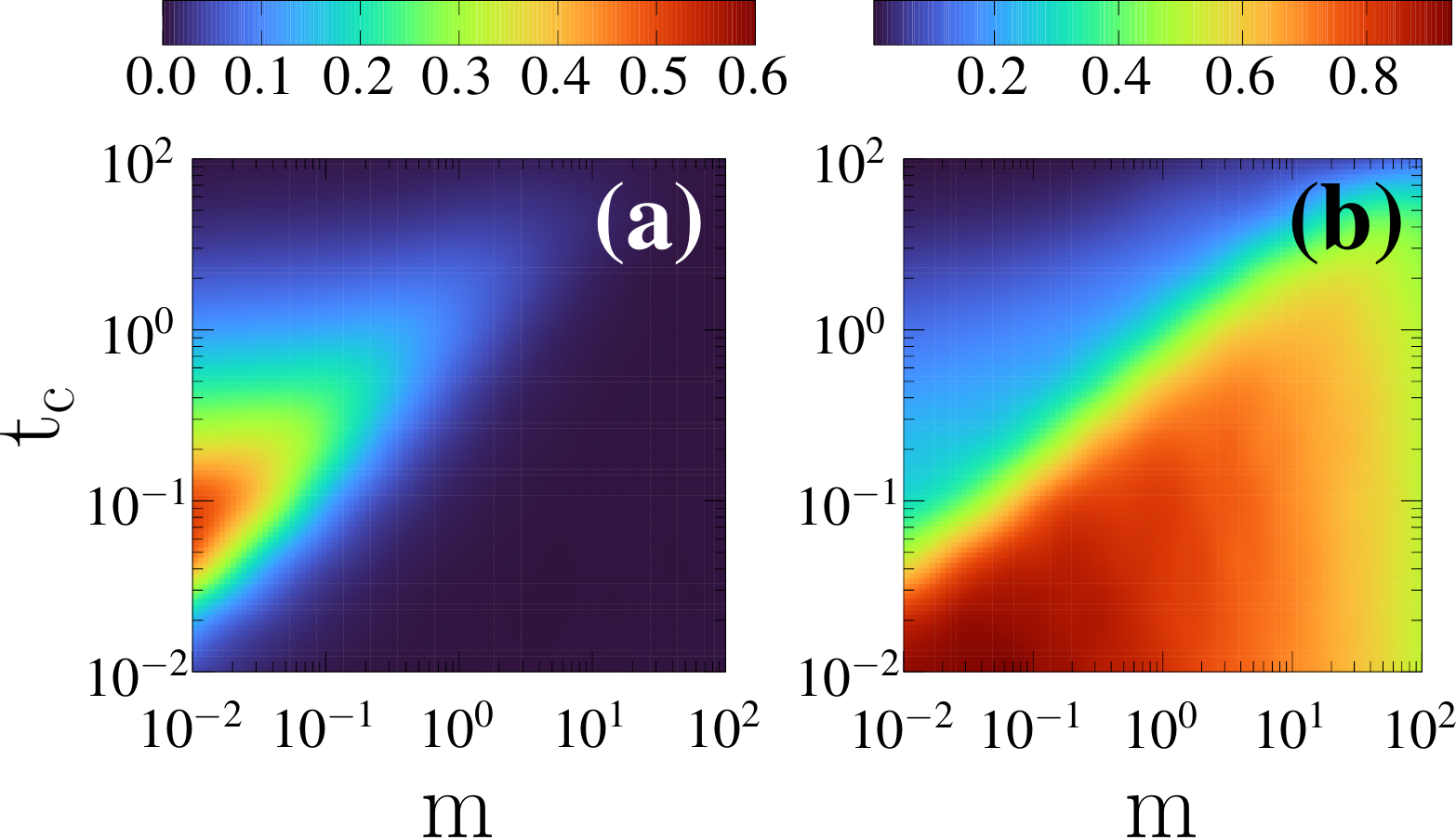}
    \caption{2D plot of $\langle j \rangle$ and $D$ as a function of $t_c$ and $m$ are shown in (a) and (b), respectively. The other common parameters are: $\Delta = 0.9\ \text{and}\ Q = 0.5$.}
    \label{fig:2d_3}
\end{figure}
\begin{figure}[!ht]
    \centering
    \includegraphics[scale=0.27]{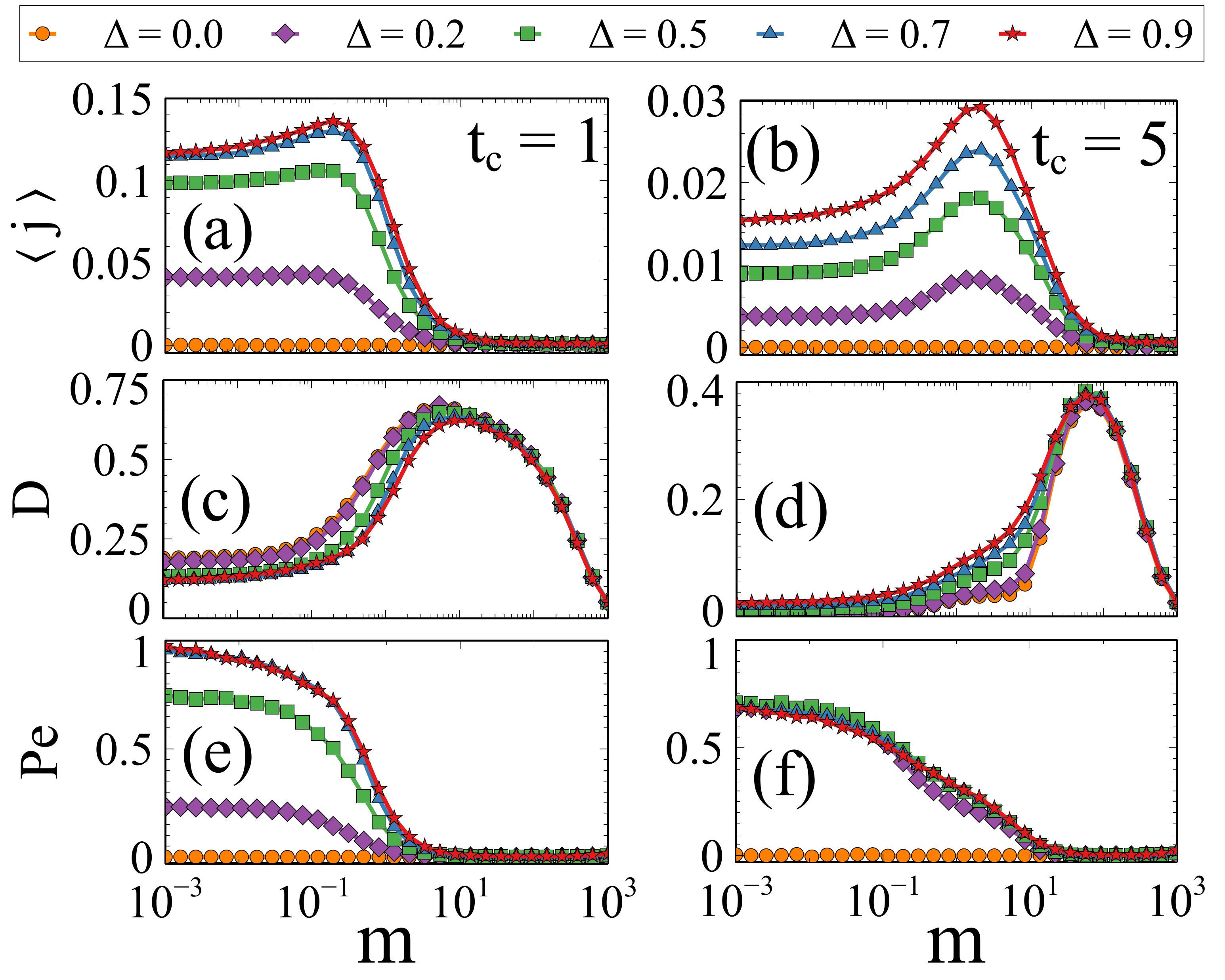}
    \caption{$\langle j \rangle$, $D$, and $Pe$ as a function of $m$ with $t_{c}=1.0$ for different values of $\Delta$ in (a), (c), and (d) and for $t_{c}=5$ in (b), (d), and (f), respectively. $Q =0.5$ is taken for all the cases.}
    \label{fig:m_for_delta}
\end{figure}

The plot of $\langle j \rangle$, $D$, and $Pe$ as a function $t_{c}$ are presented in Fig.~\ref{fig:tc_for_m}(a), (b), and (c), respectively for different values of $m$. For a given mass of the particle, $\langle j \rangle$ shows a non-monotonic behavior with $t_c$. It starts from zero and increases with $t_{c}$, attains the maximum value for an intermediate range of $t_c$, and finally approaches back to zero value for larger $t_{c}$. With increase in mass of the particle ($m$), the critical value of $t_{c}$ at which the current starts to flow, shifts towards right, reflecting that with increase in $m$, larger $t_{c}$ is required for having a net current in the ratchet. At the same time, the maximum current gets suppressed with $m$ and shifts towards larger value of $t_{c}$. This implies that for larger mass, the noise correlation in the dynamics has to persist for longer interval of time to obtain maximum current. However, $D$ shows a decaying behavior with $t_c$ as in Fig.~\ref{fig:tc_for_m}(b). For very small value of $t_c$, $D$ has maximum value which persists as long as there is no net current in ratchet. At the critical $t_{c}$, at which the current starts to flow, $D$ also decays before approaching zero as expected. It is observed that the $t_c$ at which the current has maximum value, the diffusion shows a minima type feature. Further, diffusion gets suppressed with mass of particle. As the effectiveness of the transport can be understood by analyzing the behavior of Peclet number, we have presented $Pe$ with $t_{c}$ in Fig.~\ref{fig:tc_for_m}(c). $Pe$ follows the same behavior as that of $\langle j \rangle$, confirming a coherent or reliable transport in the intermediate range of $t_{c}$. 

In order to further understand the diffusive behavior of the transport, we have simulated the mean square displacement (MSD) $\langle x^2\rangle - \langle x \rangle^2$ and plotted as a function of $t$ in Fig.~\ref{fig:tc_for_m}(d) for different values of $t_c$. For a particular $t_{c}$, in the lower time regime, MSD is found to be proportional to $t^4$, hence, the transport is super-diffusive. On the other hand, in the long time regime, the transport is diffusive in nature as the MSD is proportional to $t$. With increase in $t_{c}$, the MSD gets suppressed and approaches zero for very large $t_{c}$ values, reflecting a kind of trapping of the particle for longer persistence of noise correlation in the dynamics. To have a better understanding of the dependence of MSD with time, we introduce a parameter $\beta$ such that $\text{MSD}\propto t^\beta$. The variation of $\beta$ with time is shown in the inset of Fig.~\ref{fig:tc_for_m}(d). In the lower time regime, $\beta$ is found to be $4$, which confirms the super-diffusive transport of the particle at short timescale. In the long time limit, $\beta$ is one, which reflects the as usual steady state diffusive behavior of the particle. On the other hand, $\langle x^2 \rangle$ shows different features. In the lower time limit, it is ballistic, i.e. $\langle x^2 \rangle \propto t^2$, irrespective of the persistence duration of noise correlation in the dynamics. However, in the long time limit or at stationary state, $\langle x^2 \rangle$ depends on the correlation time. In this state, $\langle x^2 \rangle$ is diffusive (i.e. $\langle x^2 \rangle  \propto t$) for lower $t_c$ limit, ballistic (i.e. $ \langle x^2 \rangle \propto t^2$) for intermediate $t_c$ limit, and non-diffusive (i.e. independent of $t$) for larger $t_c$ limit. The different behaviors of $\langle x^2 \rangle$ and MSD in the steady state are due to the non-zero value of $\langle x\rangle$, as it is proportional to time. 
\begin{figure}[!ht]
    \centering
    \includegraphics[scale=0.387]{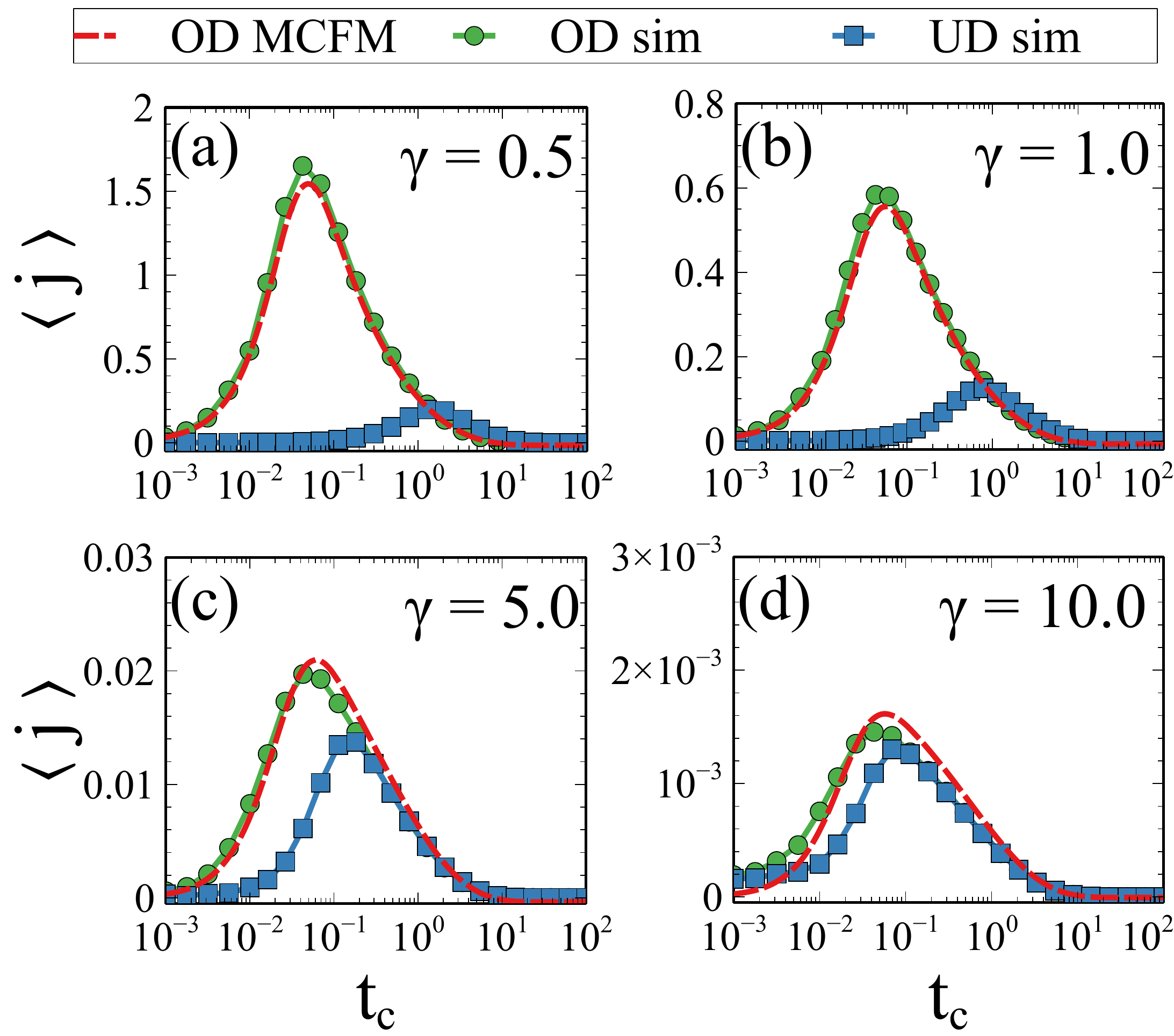}
    \caption{The simulation results of $\langle j \rangle$ as a function of $t_c$ both for overdamped (OD sim) and underdamped (UD sim) (with $m = 0.5$ ) cases along with the MCFM calculation (OD MCFM) for overdamped case are presented in (a), (b), (c), and (d) for different values of $\gamma$. The other common parameters are $\Delta = 0.9$ and $Q = 0.5$.}
    \label{fig:tc_over-under}
\end{figure}

In Fig.~\ref{fig:2d_3}(a) and (b), we depict the 2D-plots of $\langle j\rangle$ and $D$, respectively as a function of $t_c$ and $m$. It is observed that in the low $t_{c}$ limit, the current shows a monotonically decreasing behavior whereas in the intermediate regime of $t_{c}$, it shows a non-monotonic behavior with $m$. Similarly, the diffusion coefficient ($D$) shows a decreasing behavior with $m$ in the low $t_{c}$ regime and it increases with $m$ in the high $t_{c}$ regimes. In order to understand this unusual behavior of current with $m$ in the intermediate range of $t_{c}$, we have plotted $\langle j \rangle$, $D$, and $Pe$ versus $m$ for different values of $\Delta$ in Fig.~\ref{fig:m_for_delta} and each for two different values of $t_{c}$.  It is observed that current increases with $m$ and reaches a maximum value in the intermediate range of $m$ and this maximum value increases with increase in asymmetry of the potential. With further increase in $t_{c}$ value, even though the magnitude of current decreases, it shows a pronounced maximum as a function of $m$ [see Fig.~\ref{fig:m_for_delta}(b)]. $D$ shows roughly a minimum exactly around the same point, where the current shows maximum. It starts increasing as a function of $m$ from the point where current starts decreasing and finally shows a maximum at which the current approaches zero as expected. Eventhough $\langle j \rangle$ shows a prominent maximum as a function of $m$, $Pe$ does not follow the nature of $\langle j \rangle$ and it rather decreases with increase in $m$, reflecting the degradation of the coherence in the particle transport. 

Finally the simulation results of $\langle j \rangle$ as a function of $t_c$ for both the case of overdamped dynamics (excluding the inertia term) and underdamped dynamics (with inertia) for different values of $\gamma$ along with the MCFM result, are shown in Fig.~\ref{fig:tc_over-under}. It is observed that the overdamped current is always larger than that of the underdamped current as expected. With increase in $\gamma$ value, the underdamped current approaches towards the overdamped current and exactly matches with the overdamped current for very large $\gamma$ value. Further, the simulation result of average current in the overdamped limit is in good agreement with the analytical computation of average current. 

\section{SUMMARY}\label{sec:summary}
In summary, we have studied the inertial active dynamics of an Orntsein-Uhlenbeck particle in a sawtooth ratchet potential. In particular, we have investigated the transport properties as well as coherence in it's transport with the help of both analytical approximation methods and computer simulations. From the simulation results, it is inferred that potential asymmetry is the key ingredient for having a net current in the ratchet. One of our interesting findings is that the particle gets trapped in any of the minima of the ratchet potential for longer duration of self propulsion/activity in the medium, which is reflected from the simulated particle trajectories as well as from the probability distribution functions. 
In the regime of small duration of persistence of activity in the medium, the particle does not feel the influence of the potential barrier in either of the directions. As a result, it fluctuates randomly around the minimum of the potential and gets uniformly distributed through out the ratchet. In this regime, the particle behaves like an inertial passive Brownian particle that is reflected from the Gaussian velocity distribution, uniform position distribution, and the zero average current.
With further increase in the duration of self propulsion or activity in the medium, the particle starts to feel the influence of the barrier in both the directions of ratchet.
As a consequence, the free diffusion decreases and due to the asymmetry of the potential, an unidirectional net current starts to develop in the system. This can be understood from the particle trajectories, where there are intermediate abrupt jumps of the particle. 
Further, for very long duration of persistence of activity in the medium, the particle gets locked or trapped in one of the minima of the potential and randomly fluctuates across it. Because of which, both the diffusion and average particle current vanishes.

The nature of particle current is found to be non-monotonic  with the noise correlation time or self propulsion time. It increases, manifests a maximum, and then decreases as the correlation time increases. From this behavior, it is confirmed that the net transport can be controlled by fine tuning the persistent duration of activity in the medium. For an intermediate range of persistent of activity, the particle current shows a maximum as a function of mass of the particle,which is quite unusual and the absolute value of this maximum is quite sensitive to the potential asymmetry. Interestingly, it is observed that even though the particle current increases with mass in certain regime of the parameter space, the Peclet number does not follow the nature of the current and it decreases with increase in mass of the particle, confirming the degradation of reliability of transport. Moreover, we don't see any of the current reversal in our model as discussed in Refs.~\onlinecite{ai_transport_2017, xu2021rotation, reichhardt2017ratchet}. We believe that the results obtained in our model can be experimentally realized in some active matter systems in the regime of high Reynolds number. Further, it would be interesting to extend this model for investigating the collective behavior and making use of the rectified motion in terms of stochastic energetic parameters.

\section{Acknowledgement}
M.S. acknowledges the start-up grant from UGC Faculty recharge program, Govt. of India for financial support.

\end{document}